\documentclass[aps, prd, twocolumn,print, showpacs, nofootinbib]{revtex4}
\usepackage{amsmath}
\usepackage{amsfonts}
\usepackage{amssymb}
\usepackage{txfonts}
\usepackage{mathrsfs}
\usepackage{graphicx}% Include figure files
\usepackage{epsfig}
\usepackage{dcolumn}% Align table columns on decimal point
\usepackage{bm}% bold math
\begin{document}

\title{Stochastic gravitational wave background from newly born massive magnetars: The role of a dense matter equation of state}
\author{Quan Cheng$^{1\ast}$, Shuang-Nan Zhang$^1$, and Xiao-Ping Zheng$^{2,3}$}

\affiliation{$^1$Key Laboratory of Particle Astrophysics, Institute
of High Energy Physics, Chinese Academy of Sciences, Beijing 100049,
China
\\$^2$Institute of Astrophysics, Central China Normal
University, Wuhan 430079, China
\\$^3$School of Physics and Electronics Information, Hubei University of Education, Wuhan 430205, China
\\$^\ast$Electronic address: qcheng@ihep.ac.cn}

\date{Sep 2016}

\begin{abstract}
Newly born massive magnetars are generally considered to be produced
by binary neutron star (NS) mergers, which could give rise to short
gamma-ray bursts (SGRBs). The strong magnetic fields and fast
rotation of these magnetars make them promising sources for
gravitational wave (GW) detection using ground based GW
interferometers. Based on the observed masses of Galactic NS-NS
binaries, by assuming different equations of state (EOSs) of dense
matter, we investigate the stochastic gravitational wave background
(SGWB) produced by an ensemble of newly born massive magnetars. The
massive magnetar formation rate is estimated through: (i) the SGRB
formation rate (hereafter entitled as MFR1); (ii) the NS-NS merger
rate (hereafter entitled as MFR2). We find that for massive
magnetars with masses $M_{\rm mg}=2.4743 M_\odot$, if EOS CDDM2 is
assumed, the resultant SGWBs may be detected by the future Einstein
Telescope (ET) even for MFR1 with minimal local formation rate, and
for MFR2 with a local merger rate $\dot{\rho}_{\rm c}^{\rm
o}(0)\lesssim 10$ ${\rm Mpc}^{-3}{\rm Myr}^{-1}$. However, if EOS
BSk21 is assumed, the SGWB may be detectable by the ET for MFR1 with
the maximal local formation rate. Moreover, the background spectra
show cutoffs at about 350 Hz in the case of EOS BSk21, and at 124 Hz
for CDDM2, respectively. We suggest that if the cutoff at $\sim100$
Hz in the background spectrum from massive magnetars could be
detected, then the quark star EOS CDDM2 seems to be favorable.
Moreover, the EOSs, which present relatively small TOV maximum
masses, would be excluded.

\end{abstract}

\maketitle

\section{INTRODUCTION} \label{S:intro}
Short gamma-ray bursts (SGRBs) are generally considered to be
arising from the coalescence of either a neutron star-neutron star
(NS-NS) binary or a neutron star-black hole (NS-BH) binary (see
\cite{Berger:2014} for a recent review). Mergers of these compact
binaries produce strong gravitational wave (GW) emissions, making
them promising sources for GW detection using the ground based GW
interferometers such as LIGO, VIRGO, GEO600, advanced LIGO (aLIGO),
and the future Einstein Telescope (ET) \cite{Abbott:2009,Sath:2009}.
The merger product of a NS-BH binary is of course a stellar-mass BH.
On the other hand, the remnant of a NS-NS merger is still an open
question. Depending on the total mass of the binary system and the
NS equation of state (EOS), the NS-NS merger product may be either
of the following four possibilities
\cite{Rezzolla:2010,Lasky:2014,Ravi:2014,Gao:2016}: (1) a
stellar-mass BH; (2) a differential rotation supported unstable
hypermassive NS, which will collapse into a BH in a few tens
milliseconds; (3) a centrifugal force supported temporarily stable
massive NS, which will collapse into a BH when the NS is spun down;
(4) an eternally stable massive NS. The massive NS remnants are
suggested to possess strong surface dipole and internal toroidal
magnetic fields, which are amplified due to various mechanisms, such
as Kelvin-Helmholtz instability \cite{Anderson:2008},
magnetorotational instability \cite{Duez:2006} during/after the
merger, $\alpha-\omegaup$ dynamo in the nascent millisecond NS
\cite{Duncan:1992}, and the combined effect of r-mode and Tayler
instabilities \cite{Cheng:2014}. Observationally, the existence of
extended emissions \cite{Metzger:2008}, x-ray flares
\cite{Campana:2006}, and internal x-ray plateaus
\cite{Rowlinson:2013} in a large sample of SGRBs x-ray lightcurves
support the idea that the central object of some SGRBs could be a
highly magnetized, millisecond rotating NS.

Strong GW emission is expected in the final inspiral process of a
NS-NS binary. Moreover, if the merger product is either (3) or (4),
the remnant can also produce strong long-lasting GW signals, though
their strengths may be relatively weak. Generally, the fast
rotating, massive NS can emit GWs because of nonaxisymmetric
instabilities, e.g., dynamical bar-mode instability \cite{Lai:1995},
r-mode instability \cite{Andersson:1998}, f-mode instability
\cite{Andersson:1996}. On the other hand, strong internal magnetic
fields of the massive magnetar can lead to nonaxisymmetric
quadrupole deformation, which could also produce GW emission
\cite{Bonazzola:1996,Stella:2005,Dall:2009,Dall:2015}. The amplitude
of the magnetically induced GW signal is proportional to the
quadrupole ellipticity \cite{Bonazzola:1996}, which mainly depends
on the EOS, the magnetic energy, and the interior magnetic field
configuration of the NS (see, e.g.
\cite{Bonazzola:1996,Dall:2009,Gualtieri:2011,Haskell:2008,Cutler:2002,Ciolfi:2009,Mastrano:2011,Akgun:2013,Mastrano:2013,Mastrano:2015}).

Superposition of the magnetically induced GW emissions from an
ensemble of magnetars throughout the Universe can contribute to the
astrophysical stochastic GW background (SGWB). The SGWB from the
magnetic deformation of newly born magnetars has been discussed in
many literature references
\cite{Regimbau:2006,Marassi:2011,Rosado:2012,Cheng:2015}. However,
in these papers, the magnetar mass is assumed to be a canonical
value of $1.4M_\odot$, which means these magnetars are eternally
stable. Actually, the magnetars produced by NS-NS mergers apparently
have masses much larger than $1.4M_\odot$, and they may not be
always stable [e.g., product (3) mentioned above]. For a single
source, the collapse of the magnetar will lead to the cease of GW
emission at a certain frequency. Hence, in order to derive a
realistic SGWB produced by the magnetic deformation of newly born
massive magnetars, both the EOS of dense matter and the masses of
magnetars should be taken into account. In this paper, we reconsider
the SGWB produced by magnetic deformation of the newly born, massive
magnetars based on typical NS and quark star (QS) EOSs, and the
observed masses of the Galactic NS-NS binaries. The motivation of
considering the QS EOS is the recent statistical analysis of
internal x-ray plateaus of SGRBs that shows that QS remnants might
be more preferred than NSs \cite{Li:2016}. The paper is organized as
follows: in Sec. II, we show how the GW signal from a single newly
born massive magnetar is affected by the EOS. In Sec. III, we
estimate the massive magnetar formation rate (MMFR) based on two
different results: (i) the SGRB rate at redshift $z$ suggested in
\cite{Yonetoku:2014}; (ii) the NS-NS merger rate as predicted by
Regimbau and Hughes \cite{Regimbau:2009}. Results for the SGWBs
produced by an ensemble of newly born massive magnetars are shown in
Sect. IV. Conclusion and discussions are presented in Sect. V.

\section{GW EMISSION FROM THE NEWLY BORN MASSIVE MAGNETAR}\label{Sec II}
We assume that all SGRBs are produced by NS-NS mergers, and the
merger remnants are either temporarily or eternally stable massive
NSs/QSs. Generally, in the merger process, only $\lesssim
10^{-2}M_\odot$ materials are ejected from the NS-NS binary system
\cite{Hotokezaka:2013}. Hence, the total rest mass of the system is
basically conserved, e.g., $M_{\rm r,m}=M_{\rm r,1}+M_{\rm r,2}$,
where $M_{\rm r,m}$ represents the rest mass of the massive magnetar
remnant, $M_{\rm r,1}$ and $M_{\rm r,2}$ are the rest masses of the
two NSs, respectively. Based on the approximate relation
\cite{Timmes:1996} between rest and gravitational masses, by
assuming that the extragalactic NS-NS binaries have the same
gravitational mass distribution as the Galactic NS-NS binary
population, one can easily estimate the distribution of
gravitational mass $M_{\rm mg}$ for the massive magnetar remnants
\cite{Lasky:2014,Ravi:2014,Gao:2016,Li:2016}.

Until now there are six Galactic NS-NS binaries that have a
relatively accurate measured gravitational mass for each NS in the
binary system; they are PSR J0737-3039 (with gravitational masses
$M_{\rm g,1}=1.3381 M_\odot$ and $M_{\rm g,2}=1.2489 M_\odot$ for
the two NSs, respectively), PSR B1534+12 ($M_{\rm g,1}=1.3332
M_\odot$ and $M_{\rm g,2}=1.3452 M_\odot$), PSR B1913+16 ($M_{\rm
g,1}=1.4398 M_\odot$ and $M_{\rm g,2}=1.3886 M_\odot$), PSR
B2127+11C ($M_{\rm g,1}=1.358 M_\odot$ and $M_{\rm g,2}=1.354
M_\odot$), PSR J1906+0746 ($M_{\rm g,1}=1.248 M_\odot$ and $M_{\rm
g,2}=1.365 M_\odot$), and PSR J1756-2251 ($M_{\rm g,1}=1.40 M_\odot$
and $M_{\rm g,2}=1.18 M_\odot$) \cite{Kiziltan:2013}. Therefore, the
gravitational mass of the massive magnetar remnant is $2.4046
M_\odot$ as inferred from the binary system PSR J0737-3039, $2.4845
M_\odot$ inferred from PSR B1534+12, $2.6154 M_\odot$ inferred from
PSR B1913+16, $2.5139 M_\odot$ inferred from PSR B2127+11C, $2.4276
M_\odot$ inferred from PSR J1906+0746, $2.3996 M_\odot$ inferred
from PSR J1756-2251. The average gravitational mass of the massive
magnetars is thus $M_{\rm mg}=2.4743 M_\odot$, and we take this
value as the typical mass for the massive magnetars in the remainder
of this paper.

For a nonrotating NS/QS, the maximum gravitational mass that it
could sustain is the Tolman-Oppenheimer-Volkoff (TOV) maximum mass,
$M_{\rm TOV}$, which is determined by the NS/QS EOSs. However,
centrifugal forces due to the uniform rotation of the merger remnant
could increase the maximum sustainable gravitational mass. Li
\textit{et al}. \cite{Li:2016} calculated equilibrium sequences of
uniformly rotating NS/QS configurations with a spin frequency
increasing from 0 to the Keplerian spin limit and obtained
analytical expressions for the maximum gravitational mass $M_{\rm
g,max}$, the corresponding equilibrium radius $R_{\rm eq}$ (in
kilometers), and the corresponding maximum moment of inertia $I_{\rm
max}$ of a NS/QS with a spin period $P$ (in milliseconds), which,
respectively, have the following form:
\begin{eqnarray}
M_{\rm g,max}&=&M_{\rm TOV}(1+\alpha P^\beta) \label{Mgmax};\\
R_{\rm eq}&=&C+AP^B \label{Req};\\
I_{\rm max}&=&M_{\rm g,max}R_{\rm eq}^2\frac{a}{1+e^{-k(P-q)}}
\label{Imax},
\end{eqnarray}
where $M_{\rm g,max}$ and $M_{\rm TOV}$ are measured in solar
masses. The fitting parameters $\alpha$, $\beta$, $A$, $B$, $C$,
$a$, $q$, and $k$ are EOS dependent. For the typical NS (BSk21
\cite{Potekhin:2013}) and QS (CDDM2 \cite{Chu:2014}) EOSs considered
in this paper, the specific values of these parameters as well as
$M_{\rm TOV}$ can be found in Table 1 of \cite{Li:2016}.

From Eq. (\ref{Mgmax}), one can define the collapse frequency,
$\nu_{\rm coll}=1/P_{\rm coll}$, below which the massive magnetar
with a gravitational mass $M_{\rm mg}=M_{\rm g,max}(P_{\rm coll})$
will immediately collapse into a BH. The specific form of $\nu_{\rm
coll}$ is \cite{Lasky:2014,Ravi:2014,Gao:2016}
\begin{eqnarray}
\nu_{\rm coll}=\left({\alpha M_{\rm TOV} \over M_{\rm mg}-M_{\rm
TOV}}\right)^{1/\beta} \label{vcoll}.
\end{eqnarray}
If $\nu_{\rm coll}\leq0$ (i.e., $M_{\rm mg}\leq M_{\rm TOV}$), the
massive magnetar is eternally stable. However, if $0<\nu_{\rm
coll}<\nu_{\rm i}$ (with $\nu_{\rm i}$ represents the initial spin
frequency), the massive magnetar is temporarily stable, and it will
collapse into a BH when the star spins down to $\nu_{\rm coll}$ due
to GW emission and magnetic dipole radiation (MDR). Lastly, if
$\nu_{\rm coll}>\nu_{\rm i}$, the massive magnetar will collapse
into a BH immediately after it is born. Remarkable GW emissions from
the central remnant are expected only in the first two cases (i.e.,
eternally stable and temporarily stable magnetars). For the EOSs
BSk21 and CDDM2 considered here, the massive magnetar remnant with
an initial spin at the Keplerian limit should be temporarily stable
because $0<\nu_{\rm coll}<\nu_{\rm i}$ (see below). Other EOSs that
provide $M_{\rm TOV}>2.4743 M_\odot$ will result in an eternally
stable massive magnetar, and continuous GW emission extended to
lower frequencies.

The newly born massive magnetar spins down mainly through MDR and
magnetically induced GW emission. Therefore, the evolution formula
for the angular frequency $\omega$ of the magnetar can be written as
\begin{eqnarray}
\dot{\omega}=-\frac{B_{\rm
d}^2R^6\omega^3}{6Ic^3}-\frac{32G\epsilon_{\rm B}^2I\omega^5}{5c^5}
\label{dwdt},
\end{eqnarray}
where $B_{\rm d}$ is the surface dipole magnetic field at the
magnetic pole, $R$ the radius, and $I$ the moment of inertia of
star. By adopting different interior magnetic field configurations
and stellar interior structures, the magnetically induced quadrupole
ellipticity $\epsilon_{\rm B}$ has been calculated in many
literature references (e.g.,
\cite{Bonazzola:1996,Haskell:2008,Dall:2009,Dall:2015,Cutler:2002,Ciolfi:2009,Gualtieri:2011,Mastrano:2011,Akgun:2013,Mastrano:2013,Mastrano:2015}).
Some nonlinear numerical simulations show that the interior magnetic
field probably has a poloidal-toroidal ¡®twisted-torus¡¯ shape
\cite{Braithwaite:2004}. However, even for this configuration, the
dominated one is usually the toroidal field component. In the
toroidal-dominated case, $\epsilon_{\rm B}$ is related to the
volume-averaged strength of the toroidal field ${\bar B}_{\rm t}$
\cite{Cutler:2002}, which is hard to be determined directly.
Generally, ${\bar B}_{\rm t}/B_{\rm d}\approx 5$--$100$ with $B_{\rm
d}$ the dipole magnetic field of the magnetar is proposed following
the observations of giant flare from SGR 1806-20 \cite{Stella:2005},
free precession of magnetar 4U 0142+61 \cite{Makishima:2014}, x-ray
afterglows of some SGRBs \cite{Fan:2013}, and lightcurves of
superluminous supernovae \cite{Moriya:2016}. For a NS the
ellipticity can be estimated as $\epsilon_{\rm
B}\approx10^{-4}({\bar B}_{\rm t}/10^{16}~{\rm G})^2$
\cite{Cutler:2002,Lasky:2016}. While for a QS, if it is in the
two-flavor color superconductivity phase\footnote{The rotation and
temperature observations of pulsars disfavor the color-flavor-locked
QS model \cite{Madsen:2000,Cheng:2013}.} \cite{Alford:1998}, the
ellipticity is approximated as $\epsilon_{\rm
B}\approx7\times10^{-4}({\bar B}_{\rm t}/10^{16}~{\rm G})$ for the
mass and radius adopted thereinafter \cite{Glampedakis:2012}. On the
other hand, $\epsilon_{\rm B}$ can be constrained via analyzing the
internal x-ray plateau afterglows of SGRBs \cite{Gao:2016,Li:2016}.
Specifically, depending on the EOSs, the ellipticity of the massive
magnetar is confined to be $\epsilon_{\rm B}=0.002$ for NS EOS BSk21
and $\epsilon_{\rm B}=0.004$--$0.007$ for QS EOS CDDM2
\cite{Li:2016}. Thereinafter, the representative ellipticities
$\epsilon_{\rm B}=0.005$ (the value with the best Kolmogorov-Smirnov
test), and $\epsilon_{\rm B}=0.002$ will be taken for EOSs CDDM2,
and BSk21, respectively, while calculating the GW signal emitted by
a single magnetar and the SGWB from the massive magnetar population
\cite{Li:2016}. The corresponding strength of the toroidal field is
thus ${\bar B}_{\rm t}\approx 4.5\times10^{16}$ ($7.1\times10^{16}$)
G for a NS (QS).

The GW energy spectrum emitted by a single newly born massive
magnetar can be estimated as
\begin{eqnarray}
\frac{dE_{\rm GW}}{d\nu_{\rm e}}={32\pi G\over5c^5}\epsilon_{\rm
B}^2I^2\omega^6\left|\dot{\omega}^{-1}\right| \label{dEdv},
\end{eqnarray}
where $\nu_{\rm e}=\omega/\pi$ is the GW frequency at the source
frame. One can also obtain the characteristic amplitude of the
emitted GW as follows \cite{Jaranowski:1998,Corsi:2009}:
\begin{eqnarray}
h_{\rm c}(\nu_{\rm e})={\nu_{\rm e}h(t)\over\sqrt{d\nu_{\rm e}/dt}}
\label{hc},
\end{eqnarray}
where $h(t)={4\pi^2GI\epsilon_{\rm B}\nu_{\rm e}^2\over c^4d}$ is
the GW strain amplitude, $d$ is the distance to the source. To
assess the detectability of the GW signal, we calculate the optimal
(matched-filter) signal-to-noise ratio (SNR) as \cite{Corsi:2009}
\begin{eqnarray}
{\rm S/N}=\left[\int_{\nu_{\rm e,min}}^{\nu_{\rm e,max}}\frac{h_{\rm
c}^2}{\nu_{\rm e}^2S_h(\nu_{\rm e})}d\nu_{\rm e}\right]^{1/2}
\label{SNR1},
\end{eqnarray}
where $\nu_{\rm e,min}$(=$2\nu_{\rm coll}$) and $\nu_{\rm
e,max}$(=$2\nu_{\rm i}$) are, respectively, the minimum and maximum
GW frequencies emitted by the magnetar, $S_h(\nu_{\rm e})$ is the
one-sided noise power spectral density of the detector. The
analytical expressions of $S_h(\nu_{\rm e})$ can be found in
\cite{Sathyaprakash:2009} for aLIGO and ET.

We do not follow instantaneous variations of the gravitational mass,
radius, and moment of inertia with the spin-down of the massive
magnetar, though all these quantities should actually
decrease\footnote{During the spin-down process of a constant baryon
mass massive magnetar, its gravitational mass decreases more
slightly, in contrast to the radius and moment of inertia, which
show very obvious decreases (see Fig. 1 of \cite{Li:2016}).}. For
simplicity, we take a typical gravitational mass $M_{\rm mg}$,
radius $R$, and moment of inertia $I$ for the massive magnetar and
assume they do not evolve with time during spin-down. For a magnetar
with $M_{\rm mg}=2.4743 M_\odot$, its $R$ and $I$ are EOS dependent,
which can be estimated as follows. The radius $R$ is approximately
estimated by substituting the derived collapse period $P_{\rm
coll}=1/\nu_{\rm coll}$ into Eq. (\ref{Req}). Then with $R$ and
$P_{\rm coll}$, using Eq. (\ref{Imax}), the moment of inertia $I$
can be obtained approximately. The resultant radius and moment of
inertia of the $2.4743 M_\odot$ magnetar are, respectively,
$R=12.66$ (16.31) km and $I=3.68\times10^{45}$ ($5.50\times10^{45}$)
${\rm g~cm}^2$ if EOS BSk21 (CDDM2) is assumed. Obviously, $R$ and
$I$ are underestimated for the $2.4743 M_\odot$ magnetar that
initially spins at the Keplerian limit $P_{\rm K}$. As a rough
estimation, assuming a constant mass $M_{\rm mg}=2.4743 M_\odot$,
the ratio between the magnetar radii obtained at $P_{\rm K}$ and at
$P_{\rm coll}$ is $R(P_{\rm K})/R(P_{\rm coll})\approx 1.1$ (1.4)
for EOS BSk20 (CDDM1) (see Fig. 1 of \cite{Li:2016}). Furthermore,
with the spin-down of the magnetar, $R(P)/R(P_{\rm coll})$ should
decrease and become equal to 1 when $P_{\rm coll}$ is reached, where
$R(P)$ denotes the instantaneous radius of the magnetar with a spin
period $P$. For EOSs BSk21 and CDDM2 considered, $R(P_{\rm
K})/R(P_{\rm coll})$ are not expected to vary too much from the
above values. Following Eqs. (\ref{hc}) and (\ref{dEdv}), we have
$h_{\rm c}\propto R$, and $dE_{\rm GW}/d\nu_{\rm e}\propto R^2$
during the early period of spin-down when the GW emission is
dominant. Consequently, our choice of a constant $R$ will at most
underestimate the characteristic amplitude $h_{\rm c}$, and the
background emission $\Omega_{\rm GW}$ [see Eq. (\ref{omegaGW})] by a
factor of 1.4 and 2, respectively. Hence, it is reasonable to take a
constant $R$ and $I$ for a specific EOS in the calculations
below\footnote{The changes in $M_{\rm mg}$, $R$, and $I$ during the
spin-down of massive NSs are also neglected in
\cite{Ravi:2014,Gao:2016}, since these effects are unlikely to
significantly affect the evolutions of massive NSs and further their
final results.}.

\begin{figure}
\resizebox{\hsize}{!}{\includegraphics{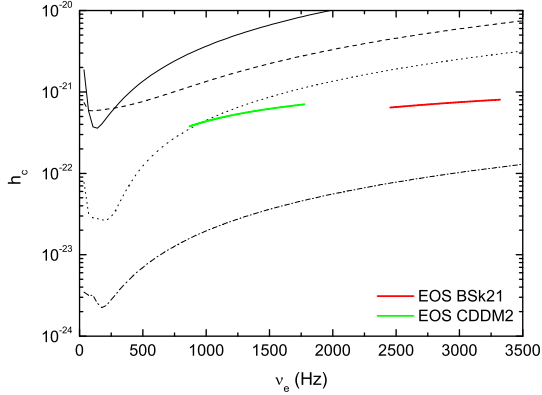}} \caption{The GW
characteristic amplitude $h_{\rm c}$ versus the emitted frequency
$\nu_{\rm e}$, calculated by assuming NS EOS BSk21 (red line) and QS
EOS CDDM2 (green line). For comparison, the rms strain noises for
LIGO (solid line), VIRGO (dashed line), aLIGO (dotted line), and the
future ET (dash-dotted line) are also presented
\cite{Sathyaprakash:2009}.} \label{Fig1}
\end{figure}

Following Li \textit{et al}. \cite{Li:2016}, the dipole magnetic
field of the massive magnetar is taken to be $B_{\rm d}=10^{15}$ G,
and the initial angular frequency is taken as $\omega_{\rm
i}=2\pi/P_{\rm K}$. The values of $P_{\rm K}$ for EOSs BSk21 and
CDDM2 can be found in \cite{Li:2016}. Assuming EOSs BSk21 and CDDM2,
we show the characteristic amplitude $h_{\rm c}$ of the GW signal
versus the emitted frequency $\nu_{\rm e}$ in Fig. \ref{Fig1}. The
distance to the source is taken to be $d=100$ Mpc. The massive
magnetar with $M_{\rm mg}=2.4743 M_\odot$ is temporarily stable for
EOSs BSk21 and CDDM2, and it will not collapse until $\nu_{\rm
coll}$ is reached. The collapse of the magnetar is manifested as a
catastrophic cutoff in the emitted GW signal at $2\nu_{\rm coll}$,
which is about 2453 Hz for EOS BSk21, and 868 Hz for EOS CDDM2, as
shown in Fig. \ref{Fig1}. For EOS BSk21, the GW signal emitted by
the massive magnetar extends from about 3322 Hz down to 2453 Hz.
While for EOS CDDM2, the emitted GW is at lower frequency band,
which covers the range from 1778 to 868 Hz.

\begin{figure}
\resizebox{\hsize}{!}{\includegraphics{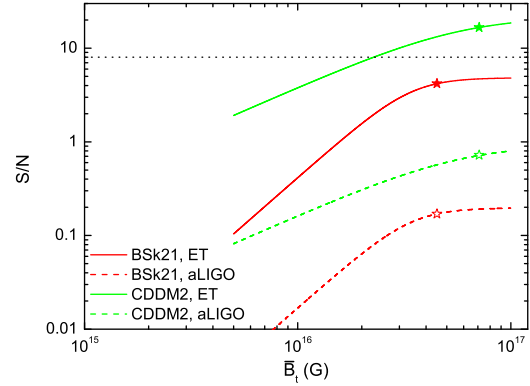}} \caption{The SNR
S/N of the GW signal emitted by a single massive magnetar versus the
magnetar's interior toroidal magnetic field ${\bar B}_{\rm t}$. The
SNRs are calculated by assuming NS EOS BSk21 (red lines), and QS EOS
CDDM2 (green lines), respectively. The solid lines represent the
SNRs obtained using future ET, while the dashed lines show the
results derived using aLIGO. The dotted line indicates the detection
threshold for a single-detector search. The stars show the SNRs
calculated by adopting various representative ellipticities and for
different detectors (see the text).} \label{Fig2}
\end{figure}

Since the strength of ${\bar B}_{\rm t}$ of a newly born magnetar is
highly uncertain, in order to comprehensively show how ${\bar
B}_{\rm t}$ could affect the detectability of GW signal from a
single source, in Fig. \ref{Fig2}, the SNR S/N is plotted as a
function of ${\bar B}_{\rm t}$, whose range is $\sim5$--$100B_{\rm
d}$ as discussed above\footnote{In newly born magnetars, ${\bar
B}_{\rm t}\sim10^{17}$ G is possible (see, e.g.
\cite{Fan:2013,Dall:2009,Cheng:2014}), since this strength is still
lower than the virial limit by about an order of magnitude
\cite{Mastrano:2011}.}. Obviously, with the increase of ${\bar
B}_{\rm t}$, the SNR is gradually enhanced. However, for different
EOSs, the evolution behaviors of S/N with ${\bar B}_{\rm t}$ differ
significantly. Compared with EOS CDDM2, as ${\bar B}_{\rm t}$
increases, S/N shows a more obvious trend of getting saturated when
EOS BSk21 is assumed. This is because for NS EOS BSk21, the
ellipticity is more sensitive to the increase of ${\bar B}_{\rm t}$
($\epsilon_{\rm B}\propto{\bar B}_{\rm t}^2$ versus $\epsilon_{\rm
B}\propto{\bar B}_{\rm t}$ for QS EOS CDDM2). When $\epsilon_{\rm
B}$ is large enough, GW emission will dominate the spin-down; thus,
we have ${\rm S/N}\propto h(t)/\sqrt{d\nu_{\rm e}/dt}={\rm const}$.
Moreover, using the same detector, the derived SNR is higher for EOS
CDDM2, since the sensitivities of the detectors are better at a
relatively low frequency band. Assuming EOS BSk21, the SNRs of the
GW emitted by the magnetar with a representative ellipticity
$\epsilon_{\rm B}=0.002$ are 4.18 for ET (red filled star in Fig.
\ref{Fig2}) and 0.17 for aLIGO (red hollow star). While assuming EOS
CDDM2 and $\epsilon_{\rm B}=0.005$, we have S/N=16.65 for ET (green
filled star in Fig. \ref{Fig2}) and S/N=0.72 for aLIGO (green hollow
star). Adopting a single-detector search, the detection threshold is
roughly S/N=8 \cite{Abadie:2010}. Hence, using ET, the emitted GW by
the $2.4743 M_\odot$ magnetar at $100$ Mpc is undetectable if EOS
BSk21 is assumed even for ${\bar B}_{\rm t}=10^{17}$ G. For
comparison, when EOS CDDM2 is assumed, a detectable GW signal is
expected if ${\bar B}_{\rm t}\gtrsim 2.3\times10^{16}$ G (see Fig.
\ref{Fig2}), which may easily be achieved for newly born magnetars.
Consequently, if future ET could detect the GW emitted by the
$2.4743 M_\odot$ magnetar, then QS EOS CDDM2 will be more preferred.
Moreover, observation of the cutoff in the GW signal using ET may
provide us an important channel to distinguish different EOSs.

\section{THE MASSIVE MAGNETAR FORMATION RATE}
As a quite rough estimation, the MMFR can be considered to be equal
to the formation rate of SGRBs because we have assumed that only
NS-NS mergers produce SGRBs and the merger products can only be
temporarily stable or eternally stable massive magnetars. Obviously,
this assumption will lead to an overestimation of the MMFR. Using
the spectral peak energy-peak luminosity correlation for SGRBs,
Yonetoku \textit{et al}. \cite{Yonetoku:2014} determined the
redshifts of 72 BATSE SGRBs and obtained the relation between the
SGRB formation rate and the redshift, which has the following form:
\begin{eqnarray}
\rho_{\rm SGRB}(z)=\left\{ \begin{aligned}
         &\rho_{\rm SGRB}(0)(1+z)^6&,~~&0<z<0.67& \\
                  &\rho_{\rm SGRB}(0)\times1.67^6&,~~&z\geq 0.67&
                          \end{aligned} \right.
                           \label{rsgrb},
\end{eqnarray}
where $\rho_{\rm SGRB}(0)$ is the local SGRB formation
rate\footnote{Since we use this formula as a rough estimation of the
MMFR, the error bars in the exponent of $(1+z)$ and the expressions
for $\rho_{\rm SGRB}(0)$ are all neglected.}. Hereafter, we refer to
this as magnetar formation rate 1 (MFR1) and assume that the MMFR
can be described by Eq. (\ref{rsgrb}) up to $z_{\ast}\sim6$. The
minimum SGRB formation rate at $z=0$ is estimated to be $\rho_{\rm
SGRB, min}(0)=1.15\times10^{-7}$ events ${\rm Mpc}^{-3}{\rm
yr}^{-1}$ by involving the geometrical correction of beaming angles
\cite{Yonetoku:2014}. On the other hand, using the peak fluxes of 14
\textit{Swift} SGRBs, redshifts, and beaming angles inferred from
x-ray observations, Coward \textit{et al}. \cite{Coward:2012}
obtained an upper limit for the local formation rate as $\rho_{\rm
SGRB,max}(0)=1.1\times10^{-6}$ events ${\rm Mpc}^{-3}{\rm yr}^{-1}$
in the case of beamed emission.

To estimate the MMFR, one can also equivalently estimate the NS-NS
merger rate under the assumption that NS-NS mergers can only produce
temporarily stable or eternally stable massive magnetars. Hereafter,
we refer to the MMFR derived in this way as magnetar formation rate
2 (MFR2). Assuming that the NS-NS merger rate tracks the cosmic star
formation rate (CSFR) with the time delay $t_{\rm d}$ from formation
of the NS binary to the final merger, the observed NS-NS merger rate
at redshift $z$ can be written as \cite{Regimbau:2009}
\begin{eqnarray}
\dot{\rho}_{\rm c}^{\rm o}(z)=\dot{\rho}_{\rm c}^{\rm
o}(0)\times\frac{\dot{\rho}_{\rm \ast,c}(z)}{\dot{\rho}_{\rm
\ast,c}(0)} \label{rhocz},
\end{eqnarray}
where $\dot{\rho}_{\rm c}^{\rm o}(0)$ is the observed local merger
rate per unit volume, whose value can be extrapolated by multiplying
the Galactic NS-NS merger rate with the density of Milky Way-like
galaxies. Following Regimbau and Hughes \cite{Regimbau:2009}, we
take two representative values for the local merger rate: (i)
$\dot{\rho}_{\rm c}^{\rm o}(0)=10$ ${\rm Mpc}^{-3}{\rm Myr}^{-1}$,
which is the upper limit of the local merger rate; (ii)
$\dot{\rho}_{\rm c}^{\rm o}(0)=1$ ${\rm Mpc}^{-3}{\rm Myr}^{-1}$,
which represents the most probable value for the rate. The NS-NS
merger rate is related to the CSFR by the quantity $\dot{\rho}_{\rm
\ast,c}(z)$, which can be given as \cite{Regimbau:2009}
\begin{eqnarray}
\dot{\rho}_{\rm \ast,c}(z)=\int\frac{\dot{\rho}_{\rm \ast}(z_{\rm
f})}{1+z_{\rm f}}P(t_{\rm d})dt_{\rm d},
\end{eqnarray}
where $\dot{\rho}_{\rm \ast}$ is the CSFR. $z$ and $z_{\rm f}$ are
the redshifts at which the NS-NS binary mergers and its progenitor
binary initially formed, respectively. $t_{\rm d}$ is the time
difference between the formation of the progenitor binary and the
compact binary, plus the merging time of the binary. It also
represents the lookback time between $z$ and $z_{\rm f}$, which has
an approximate form $t_{\rm d}\simeq{2\over3H_{\rm
0}}(1/\sqrt{\Omega_m(1+z)^3+\Omega_\Lambda}-1/\sqrt{\Omega_m(1+z_{\rm
f})^3+\Omega_\Lambda})$. In this paper, the $\Lambda$CDM
cosmological model is taken with the Hubble constant $H_{\rm
0}=70~{\rm km~s}^{-1}{\rm Mpc}^{-1}$, $\Omega_m=0.3$, and
$\Omega_\Lambda=0.7$. Based on the result of population synthesis
(see \cite{Regimbau:2009} and references therein), the probability
distribution $P(t_{\rm d})$ for the time delay $t_{\rm d}$ is given
by
\begin{eqnarray}
P(t_{\rm d})\propto1/t_{\rm d}~~{\rm with}~~t_{\rm d}>\tau_{0},
\end{eqnarray}
for some minimal delay time $\tau_{0}$. For NS-NS binary, the
minimal delay time is assumed to be $\tau_{0}=20$ Myr, which
corresponds to the evolution time from massive binaries to NS-NS
binaries \cite{Regimbau:2009}.

For the CSFR, we use the result suggested in Hopkins and Beacom
\cite{Hopkins:2006}. Based on the new measurements of the galaxy
luminosity function in the UV and far-infrared wavelengths, they
refined the previous models up to $z_{\ast}\sim6$ and obtained a
parametric fit formula for the CSFR, which takes the following form
\cite{Hopkins:2006}:
\begin{eqnarray}
\dot{\rho}_{\rm \ast}(z)=h\frac{0.017+0.13z}{1+(z/3.3)^{5.3}}M_\odot
{\rm yr}^{-1}{\rm Mpc}^{-3},
\end{eqnarray}
where $h=0.7$.

\section{RESULTS}
The SGWB is generally represented by the dimensionless quantity,
$\Omega_{\rm GW}(\nu_{\rm obs})$, which describes the distribution
of the GW energy density versus the GW frequency in the observer
frame $\nu_{\rm obs}$. The SGWB produced by the magnetic deformation
of an ensemble of newly born massive magnetars is given by
\cite{Regimbau:2006,Marassi:2011,Cheng:2015}
\begin{eqnarray}
\Omega_{\rm GW}(\nu_{\rm obs})=\frac{8\pi G\nu_{\rm
obs}}{3H_0^3c^2}\int_{z_{\rm low}}^{z_{\rm upp}}\frac{\rho_{\rm
MFR}(z)}{(1+z)E(\Omega,z)}\frac{d E_{\rm GW}}{d \nu_{\rm
e}}dz,
\label{omegaGW}
\end{eqnarray}
where $\nu_{\rm obs}=\nu_{\rm e}/(1+z)$,
$E(\Omega,z)=\sqrt{\Omega_m(1+z)^3+\Omega_\Lambda}$, and $\rho_{\rm
MFR}(z)$ is the MMFR, which can be substituted by Eqs. (\ref{rsgrb})
or (\ref{rhocz}). $z_{\rm upp}$ and the $z_{\rm low}$ are the upper
and lower limits of the redshift integration, respectively. $z_{\rm
upp}$ depends on the maximal redshift $z_{\ast}$ of the MMFR model
and the maximal value of $\nu_{\rm e}$, i.e., $z_{\rm
upp}=$min$(z_{\ast}, \nu_{\rm e,max}/\nu_{\rm obs}-1)$. While
$z_{\rm low}$ is determined by the minimal value of $\nu_{\rm e}$,
i.e., $z_{\rm low}=$max$(0, \nu_{\rm e,min}/\nu_{\rm obs}-1)$.

The optimal SNR of the background emission for an observation time
$T$ is given as \cite{Allen:1999}
\begin{eqnarray}
({\rm S/N})_{\rm
B}=\left[\frac{9H_0^4T}{50\pi^4}\int_0^\infty\frac{\gamma^2(\nu_{\rm
obs})\Omega_{\rm GW}^2(\nu_{\rm obs})}{\nu_{\rm
obs}^6S_{h1}(\nu_{\rm obs})S_{h2}(\nu_{\rm obs})}d\nu_{\rm
obs}\right]^{1/2}\label{SNR1},
\end{eqnarray}
where $S_{h1}(\nu_{\rm obs})$, $S_{h2}(\nu_{\rm obs})$ are the noise
power spectral densities of the two detectors, and $\gamma$ is the
normalized overlap reduction function. For two colocated and
coaligned detectors, $\gamma=1$, and we simply assume
$S_{h1}(\nu_{\rm obs})=S_{h2}(\nu_{\rm obs})$.
\begin{figure}
\resizebox{\hsize}{!}{\includegraphics{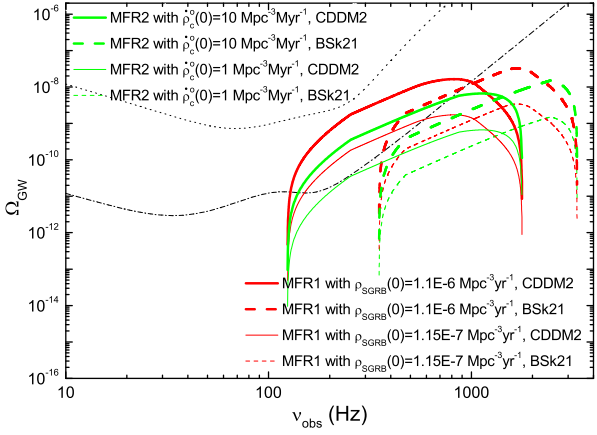}}
\caption{Dimensionless GW energy density $\Omega_{\rm GW}$ versus
the observational frequency $\nu_{\rm obs}$, calculated for
different MMFR models with various local formation rates and
different EOSs as shown in the legend. The newly born massive
magnetars have $M_{\rm mg}=2.4743 M_\odot$, and $B_{\rm d}=10^{15}$
G. Other quantities for the magnetars are taken to be EOS dependent
(see the text). For comparison, the detection thresholds of aLIGO
(black dotted line) and the future ET (black dash-dotted line)
calculated following Eq. (136) in \cite{Sathyaprakash:2009} and
assuming a 1 yr observation time are also shown.} \label{Fig3}
\end{figure}

By considering different MMFR models (MFR1 and MFR2) and EOSs (BSk21
and CDDM2), in Fig. \ref{Fig3}, we plot the SGWBs contributed by an
ensemble of newly born massive magnetars with $M_{\rm mg}=2.4743
M_\odot$. As mentioned before, depending on the EOS, the radii,
representative ellipticities, and initial spin frequencies of
magnetars are taken to be $R=12.66$ (16.31) km, $\epsilon_{\rm
B}=0.002$ (0.005), and $\nu_{\rm i}=1660.85$ (888.97) Hz for EOS
BSk21 (CDDM2), while the dipole magnetic fields are taken the same
as $B_{\rm d}=10^{15}$ G for all magnetars \cite{Li:2016}.
Consequently, the maximal observed GW frequency is $\nu_{\rm
obs}^{\rm max}\simeq3322$ Hz for EOS BSk21 and 1778 Hz for EOS
CDDM2. For the same EOS, the background spectra calculated by using
MFR1 have different shapes in comparison with those derived by using
MFR2. For instance, the peak frequencies are at 830$-$1690 Hz for
MFR1 versus 1132$-$2455 Hz for MFR2. Moreover, assuming EOS CDDM2,
the spectrum calculated based on MFR1 with $\rho_{\rm
SGRB}(0)=1.1\times10^{-6}$ events ${\rm Mpc}^{-3}{\rm yr}^{-1}$ (red
thick solid line) dominates the spectrum obtained by using MFR2 with
$\dot{\rho}_{\rm c}^{\rm o}(0)=10$ ${\rm Mpc}^{-3}{\rm Myr}^{-1}$
(green thick solid line) at $\nu_{\rm obs}\lesssim 1200$ Hz,
however, succumbs to the later above 1200 Hz. This is because MFR1
predicts higher $\rho_{\rm MFR}(z)$ at high $z$, but lower
$\rho_{\rm MFR}(z)$ at low $z$. The source formation rate at high
(low) $z$ mainly contributes to background emission at low (high)
frequency band \cite{Zhu:2011}.

\begin{figure}
\resizebox{\hsize}{!}{\includegraphics{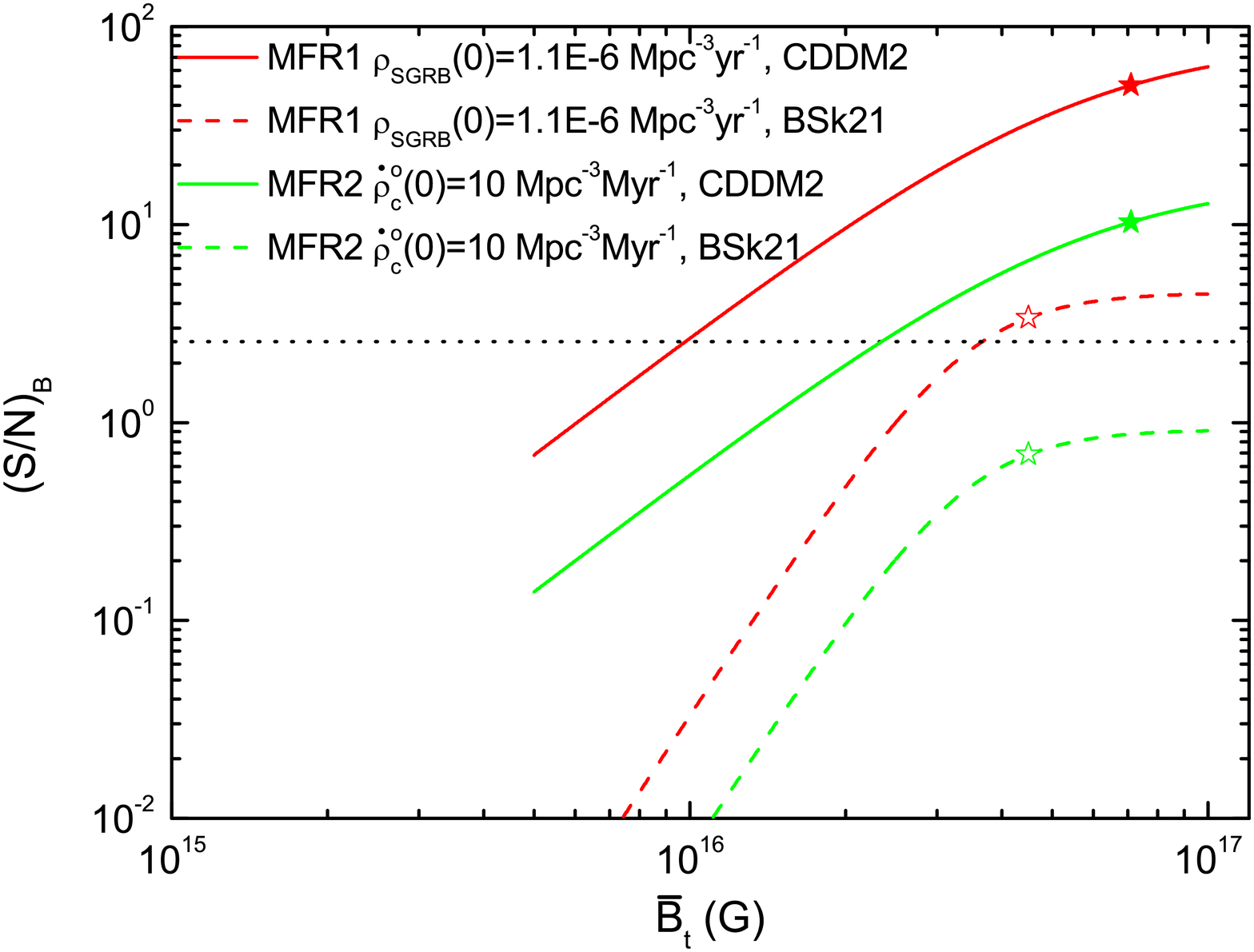}} \caption{The SNR
$(\rm S/N)_{\rm B}$ of SGWB from a massive magnetar population
versus the magnetars' interior toroidal magnetic fields ${\bar
B}_{\rm t}$. The SNRs are calculated by assuming different EOSs and
various MMFRs as depicted in the legend. All SNRs are derived with
respect to ET for an observation time of 1 yr. The dotted line shows
the detection threshold of ET. The stars represent the SNRs
calculated by adopting various representative ellipticities
(corresponding to various EOSs) and for different MMFRs (see the
text).} \label{Fig4}
\end{figure}

It is obvious that the background spectra are strongly dependent on
the assumed EOSs as shown in Fig. \ref{Fig3}. The EOS BSk21 leads to
a cutoff at $\nu_{\rm e}\simeq 2453$ Hz in the GW signal emitted by
a single magnetar with $M_{\rm mg}=2.4743 M_\odot$ (see Fig.
\ref{Fig1}). Such cutoffs also appear in the background spectra
emitted by an ensemble of such magnetars if EOS BSk21 is assumed.
However, the observed cutoff frequency depends on the maximal
redshift of the MMFR model as $\nu_{\rm
cut}=2453/(z_{\ast}+1)\simeq350$ Hz with $z_{\ast}=6$ for the two
MMFR models. In contrast, using a commonly assumed mass $1.4M_\odot$
for newly born magnetars, the resultant background spectrum extends
down to several hertz without a cutoff because these magnetars do
not collapse
\cite{Regimbau:2006,Marassi:2011,Rosado:2012,Cheng:2015}. With EOS
BSk21, the background emissions cover the frequency band from 2453
to 350 Hz, which are not in the sensitive band of ET. The background
spectrum may only be detected by ET in the case of MFR1 with
$\rho_{\rm SGRB}(0)=1.1\times10^{-6}$ events ${\rm Mpc}^{-3}{\rm
yr}^{-1}$ (red thick dashed line in Fig. \ref{Fig3}). The
corresponding SNR of the background spectrum is $(\rm S/N)_{\rm
B}=3.38$ (red hollow star in Fig. \ref{Fig4}) for ET when an
observation time $T=1$ yr is assumed. The SNR is slightly above the
detection threshold $(\rm S/N)_{\rm B, th}=2.56$ of ET (dotted line
in Fig. \ref{Fig4}) \cite{Marassi:2011}. A lower SNR with $(\rm
S/N)_{\rm B}=0.69$ (green hollow star in Fig. \ref{Fig4}) is
obtained using MFR2 with $\dot{\rho}_{\rm c}^{\rm o}(0)=10$ ${\rm
Mpc}^{-3}{\rm Myr}^{-1}$. Consequently, it should be hard to confirm
EOS BSk21 through direct observation of such a cutoff in the
background spectrum emitted by massive magnetars.

As a comparison, if the EOS CDDM2 is assumed, the resultant
background spectra emitted by all magnetars with $M_{\rm mg}=2.4743
M_\odot$ show cutoffs at $\nu_{\rm cut}\simeq124$ Hz. The spectra
extend from 1778 Hz down to 124 Hz. Using ET and taking $T=1$ yr,
the SNR of the background spectrum is 50.49 (red filled star in Fig.
\ref{Fig4}) for MFR1 with $\rho_{\rm SGRB}(0)=1.1\times10^{-6}$
events ${\rm Mpc}^{-3}{\rm yr}^{-1}$. The SNR is reduced by about an
order of magnitude for the minimal local SGRB formation rate
$\rho_{\rm SGRB, min}(0)=1.15\times10^{-7}$ events ${\rm
Mpc}^{-3}{\rm yr}^{-1}$. Moreover, adopting MFR2 with
$\dot{\rho}_{\rm c}^{\rm o}(0)=10$ ${\rm Mpc}^{-3}{\rm Myr}^{-1}$,
the resultant SNR is 10.28 for ET (green filled star in Fig.
\ref{Fig4}). The SNRs are all above the detection threshold of ET,
which suggest that the spectra may be detectable by the proposed ET.
If the cutoff at $\sim100$ Hz in the SGWB from massive magnetars
could be detected in the future, then the EOS of dense matter may be
consistent with QS EOS CDDM2. Furthermore, the EOSs (e.g., BSk21 and
APR) which provide relatively small $M_{\rm TOV}$ could be excluded.
Of course, the nondetection of the background emission may have the
following reasons: (i) the actual $\rho_{\rm MFR}(z)$ is much
smaller than those we adopted here; (ii) the EOSs (e.g., BSk21 and
APR) which provide relatively small $M_{\rm TOV}$ are favorable;
(iii) the magnetars actually have much lower ${\bar B}_{\rm t}$.

Related to the last point of the above reasons, the effect of ${\bar
B}_{\rm t}$ on the detection of the SGWB is shown in Fig.
\ref{Fig4}. All the SNRs of the background emissions are calculated
by adopting the maximal local event rate of each MMFR model, an
observation time $T=1$ yr, and with respect to ET, which has the
best designed sensitivity. In all cases, as ${\bar B}_{\rm t}$
increases, the SNR of the background first rises rapidly then
slowly. Specially, the same as in Fig. \ref{Fig2}, when EOS BSk21 is
assumed the SNR of the background tends to be saturated when ${\bar
B}_{\rm t}$ becomes large enough. In this case, even MFR1 with
$\rho_{\rm SGRB}(0)=1.1\times10^{-6}$ events ${\rm Mpc}^{-3}{\rm
yr}^{-1}$ and the maximal toroidal fields ${\bar B}_{\rm
t}\sim10^{17}$ G are taken, $(\rm S/N)_{\rm B}$ is slightly above
detection threshold of ET. This just reflects that the background
emission from massive magnetars is difficult to be detected if EOSs
with small $M_{\rm TOV}$ are preferred as discussed above. In
contrast, for the same MMFR and ${\bar B}_{\rm t}$, $(\rm S/N)_{\rm
B}$ derived based on EOS CDDM2 is at least $\sim10$ times higher
than that obtained by assuming EOS BSk21 (see Fig. \ref{Fig4}).
Hence, if EOS CDDM2 rather than EOS BSk21 is preferred, detection of
the background emission from massive magnetars may be promising.

The SGWB produced by an ensemble of massive magnetars is expected to
be continuous if the duty cycle $DC=\int_0^6\tau(1+z)\rho_{\rm
MFR}(z)dV\gg 1$ is satisfied \cite{Coward:2006}. The quantity $\tau$
is the duration of the GW signal emitted by a single magnetar. The
expression for the comoving volume element $dV$ can be found in
\cite{Zhu:2011}. In the case of EOS BSk21, since the $2.4743M_\odot$
massive magnetar has a dipole field $B_{\rm d}=10^{15}$ G and
representative ellipticity $\epsilon_{\rm B}=0.002$, its lifetime
can be determined to be $\tau=140.50$ s. The massive magnetar has a
much longer lifetime of $\tau=995.86$ s if EOS CDDM2 is assumed,
even though its representative ellipticity ($\epsilon_{\rm
B}=0.005$) is larger in this case. Using MFR2 with $\dot{\rho}_{\rm
c}^{\rm o}(0)=1$ ${\rm Mpc}^{-3}{\rm Myr}^{-1}$ (the lowest MMFR),
we have $DC\simeq 39.02$ (276.56) for EOS BSk21 (CDDM2), which means
that the SGWB from these magnetars is continuous. We note that even
when EOS BSk21 and ultrastrong toroidal fields ${\bar B}_{\rm
t}\sim10^{17}$ G are assumed for the massive magnetars, the produced
SGWB is still continuous if the MMFR is not as low as MFR2 with
$\dot{\rho}_{\rm c}^{\rm o}(0)=1$ ${\rm Mpc}^{-3}{\rm Myr}^{-1}$.

\section{CONCLUSION AND DISCUSSIONS}\label{Sec V}
As one of the most promising targets for GW detection, newly born
magnetars, if produced by NS-NS mergers, their masses should be much
larger than the generally assumed value $1.4 M_\odot$ for NSs. The
masses of the newly born magnetars, the EOSs, and the MMFR models
all have an impact on the SGWB produced by the massive magnetar
population. By taking into account these effects, we estimated the
SGWB produced by the newly born massive magnetars. For the NS EOS
BSk21 and QS EOS CDDM2 adopted here, the resultant background
spectra contributed by massive magnetars with $M_{\rm mg}=2.4743
M_\odot$ show cutoffs at about 350 Hz, and 124 Hz, respectively. The
frequency ranges of background emissions are different for the two
EOSs. Assuming EOS CDDM2 and representative ellipticities
$\epsilon_{\rm B}=0.005$ for the $2.4743 M_\odot$ massive magnetars,
even using MFR1 with the minimal local rate, the SGWB contributed by
an ensemble of such magnetars may be detected by the future ET.
While using MFR2, the background emission from these magnetars may
be detected by ET only if the local merger rate satisfies
$\dot{\rho}_{\rm c}^{\rm o}(0)\lesssim 10$ ${\rm Mpc}^{-3}{\rm
Myr}^{-1}$. However, if EOS BSk21 (and representative ellipticities
$\epsilon_{\rm B}=0.002$) is assumed, the SGWB may be detected by ET
only when MFR1 with the maximal local rate is adopted. For the same
MMFR and ${\bar B}_{\rm t}$, adopting EOS CDDM2, the SNR of the
background is at least $\sim10$ times higher than that obtained
based on EOS BSk21. This, in turn, may indicate that if such
background emission could be detected, EOS CDDM2 should be more
favorable. The relatively low SNR of the background emission
indicates that it may be unlikely to test EOS BSk21 via detecting
the cutoff in the background spectrum. However, if the cutoff at
$\sim100$ Hz in the SGWB from massive magnetars could be detected in
the future, the QS EOS CDDM2 seems to be favorable. In addition,
successful detection of background emission at $\gtrsim100$ Hz could
reasonably exclude the EOSs which present relatively small $M_{\rm
TOV}$ (e.g., EOSs BSk21 and APR). Finally, detecting the GW emission
during the entire formation process (from the final binary inspiral
process to the magnetar phase) of a newly born massive magnetar may
still be an effective way to probe the EOS of dense matter because
one need not estimate the MMFR, which is actually rather uncertain.

Improvements are still needed in order to obtain a more realistic
SGWB produced by an ensemble of massive magnetars. In this paper, we
only assume typical mass $M_{\rm mg}=2.4743 M_\odot$ for the massive
magnetars. Actually, the masses of these massive magnetars should
distribute in a certain range as inferred from the mass distribution
of Galactic NS-NS binaries \cite{Kiziltan:2013}. In future work, we
will combine the mass distribution of massive magnetars with various
NS/QS EOSs to study the SGWB produced by the massive magnetars in
detail.

\acknowledgements We gratefully thank the anonymous referee for
insightful comments and suggestions in improving this paper. We also
thank Y. W. Yu, X. L. Fan, and H. Gao for helpful discussions. This
work is supported by the National Natural Science Foundation of
China (Grants No. 11133002 and No. 11178001).

\end{document}